\documentclass[journal]{IEEEtran}
\usepackage[cmex10]{amsmath}
\usepackage{flushend}
\usepackage{cite}
\usepackage{url}
\usepackage{ifpdf}
\ifCLASSINFOpdf
  \usepackage[pdftex]{graphicx}
  \DeclareGraphicsExtensions{.pdf}
\else
  \usepackage[dvips]{graphicx}
   \DeclareGraphicsExtensions{.eps}
\fi
\begin{document}
\title{Search in the Universe of Big Networks and Data}
\author{Erol~Gelenbe,~\IEEEmembership{Fellow,~IEEE,}
        and~Omer~H.~Abdelrahman,~\IEEEmembership{Member,~IEEE}%
\thanks{Manuscript received December 10, 2013; revised April 4, 2014; accepted April 15, 2014.}
\thanks{E. Gelenbe and O. H. Abdelrahman are with the Department
of Electrical and Electronic Engineering, Imperial College, London SW7 2BT, UK, email: \{e.gelenbe,o.abd06\}@imperial.ac.uk.}}
\markboth{IEEE Network,~Vol.~XX, No.~X, MONTH~2014}{}
\maketitle
\begin{abstract}
Searching in the Internet for some object characterised by its attributes in the form of data, such as a hotel in a certain city whose price is less than something, is one of our most common activities when we access the Web. We discuss this problem in a general setting, and compute the average amount of time and the energy it takes to find an object
in an infinitely large search space. We consider the use of $N$ search agents which act concurrently. Both
the case where the search agent knows which way it needs to go to find the object, and the case where the search agent is perfectly ignorant and may even head
away from the object being sought. We show that under mild conditions regarding the randomness of the search and the use of a time-out, the search agent will always find the object despite the fact that the search space is infinite. We obtain a formula for the average search time and the average energy expended by $N$ search agents acting concurrently and independently of each other. We see that the time-out itself can be used to minimise the search time and the amount of energy that is consumed to find an object. An approximate formula is derived for the number of search agents that can help us guarantee that an object is found in a  given time, and we discuss how
the competition between search agents and other agents that try to hide the data object, can be used by opposing parties to guarantee their own success.
\end{abstract}

\section{Introduction}\label{Basic}

The conventional view that the Internet is very large but finite follows from the fact that the set of all possible Internet addresses is finite,
and in addition that the memory of any computer connected to the Internet is also finite.
Though this view may dominate  for many more years,
the world of networks and data is so large that this is not  a convenient way to represent reality.
Thus our paper takes a continuous and infinite view of such systems, in line with approaches taken in other fields of science with regard to search for specific objects.
In such systems we are faced with three basic questions:
\begin{enumerate}
\item Can we estimate the time it takes to find a particular object?
\item Is there any certainty that we will actually find the data object in finite time? Or will we just engage in an eternal and futile search?
\item Since any search activity is bounded by key resources such as computational time and, increasingly by energy consumption, is there a way to estimate whether objects may be found with finite energy and if so what is the energy needed?
\end{enumerate}
Calling upon recent research, this paper will attempt to answer some of these questions.

\subsection{Prior Work} \label{Prior}

Just as in molecular science the identity and location of the molecules with which a given ``searcher molecule'' will bind are unknown, in network and data engineering we often do not know the characteristics of the data items which are of interest to us, nor do we know which specific  data items actually respond to
the requirements of a search or where they are.
Only search in an unbounded data and network universe will eventually provide some data items that
are of interest to the end user.

Within computer science, many efforts have been devoted to developing appropriate languages and data organisations that can help formulate
requests, organise the data appropriately in advance of search to achieve fast data identification and extraction from the web \cite{Furche,DBLP2012}. In packet and peer-to-peer networks,
effective algorithms have been studied for neighbourhood search \cite{Elias,Discrete}, and routing in opportunistic networks \cite{Wang05} while diffusion models have also been studied for sensor networks \cite{Shakkottai05}.

Learning based search in large networks \cite{SAN} is also studied for large Clouds
where network and data links \cite{Misbah} make the network connexions more complex.
Search for objects which are hard to find has been addressed in robotics for dangerous environments \cite{Search-mines,Mines,CAMWA2012} but
this still remains an area that relies heavily on human guidance and intuition.

Much of the earlier theoretical work on search originates in biology \cite{Gelenbe1997,Sims08}. Physics  \cite{Benichou2004} and biochemistry \cite{Eliazar2008} often consider  ``an infinite universe for search'', where the objects (molecules) that conduct the search
and the objects that need to be found (yet other molecules with which they may react or bind, or specific sites on cells) are tiny  \cite{Tilch99a}.  Moving or fixed objects are a few angstroms in size, and move in
3D volumes which are millimeters, centimetres or larger in diameter. Furthermore the actual boundaries of the volumes considered may
be difficult to define. Similar problems arise in biology where predators seek preys in a very large search space \cite{Oshanin2009}.
Much of the work in this area uses diffusion models which  offer a
relatively simple and convenient analytical framework to represent approximately the discrete time steps of a long search process,
together with {\em L\'evy flights} or instantaneous jumps \cite{Redner1}  to represent the accelerated movement of the searcher, or the sudden escape of the object that is being sought.
Such models were also used for performance analysis of computer systems. For instance, when a diffusion represents  the waiting time for a processing unit or a memory sub-system, the instantaneous jump is  the service time of the first request that arrives when the processor or memory unit is idle \cite{Gelenbe1976,Czachorski08}.

The situation is similar in very large networks containing huge volumes of data \cite{Saarland}. The objects which conduct the search are
programs which crawl through the web, looking for specific data objects which are  tiny
in comparison to the vast data sets contained in  thousands of computers.

\subsection{Contents of the Paper}

The paper is organised as follows. Section \ref{model} discusses a framework within which the above questions may be addressed using a simple mathematical model that represents a finite set of $N$ statistically identical searchers that seek a given object independently of each other. The  searchers are subject to ``accidents''  and
may be destroyed or ``killed''. However each of the searchers may search  ``for ever'' if they are left to their own devices, simply because the search space is infinite. Thus, a time-out is set for each of the searchers independently, and when an individual searcher's time-out runs out, the searcher
stops in its tracks and is removed and it will
(after another finite time interval) be replaced by a new searcher that acts independently but statistically identically, with respect to its previous incarnation. In the sequel we will see that the time-out is essential, and that together with the statistically identical behaviour of successive incarnations of searchers, it guarantees that the object will eventually be found even when $N=1$.

In Section \ref{avg-comput} we  see that even when searchers make poor choices
about the direction in which they should orient their search, and
drift progressively away from the object being sought, one of the searchers will eventually find the object even when $N=1$. The formula for the average search time also allows us to estimate how the search time is reduced as a function of $N$.
We then consider the case where the success of the search requires that out of the $N$ searchers, $k$ of them must be successful.
This arises in different applications. For instance,  if multiple searchers all confirm that an object is found, this will provide greater assurance that the object  has actually been found. Also, once an object is found it may be necessary for multiple agents to provide separate keys
or process it in order to extract its content.
When coded packets are transmitted, it may be necessary to receive or find multiple copies of the
packet in order to decode it correctly \cite{Wang05}.

Thus Section \ref{asymptot1} presents a
formula that is valid for any large value of time $B$. This formula provides a quick estimate of the number $N$ of searchers which are needed if $k$ of them
are required to find the object within time $B$.

While time is an important factor, an increasingly important  and critical resource for ICT  is energy consumption. Thus the amount
of energy consumed by $N$ searchers during a successful search is also an important issue that we discuss in Section \ref{energy}.

The first part of the analysis assumes that the search occurs in a spatially homogeneous and time-invariant
search space. Yet search in non-homogeneous media is quite important in many cases.
For instance, in security applications, if a malignant searcher is trying to find a specific web site or network node that it wishes to attack, that site or node may
in turn be much better protected in its own immediate vicinity so that the malignant search agent will have a harder time finding its target as it gets closer to it.
This is the case when, for instance, a network node is being defended against Cyberattacks, and
deep inspection of packets approaching the node are being carried out and suspicious looking packets are simply dropped.

Thus  in Section \ref{non-hom} we describe results based on a numerical approach that computes the search time and the energy needed to find an object when the search space is non-homogeneous. In particular,
we show that a form of phase transition occurs when the
rate at which searchers are lost or destroyed is adjusted in conjunction with the speed at which the
search agent can move towards the data object that it is seeking. With a proper choice of its speed, the searcher will always be successful in finding a well concealed or protected item,
while in other cases it will never be successful in finding the object.

\section{Modelling the Search Process}\label{model}

Consider $N$ independent search agents that are sent out simultaneously at time $t=0$, in the quest for the same object. Let $Z_i(t)$ be the $i$-th searcher's {\em distance} from its destination at time $t\geq 0$. Since all the searchers start from the same point, their unknown distance from that object when they start is $D$ and therefore  $Z_i(0)=D$.

The time it takes the $i$-th searcher to  find the object is simply the first time that this distance $Z_i(t)$ becomes zero, or $T_i = \inf\{t: Z_i(t)=0 \}$. Take the $N$ values $T_1,~...~T_N$ and order them so that $T_{1,N}\leq T_{2,N}\leq \cdots \leq T_{N,N}$ are  the variables $T_i$ rearranged in ascending order.

\begin{figure}[t]\centering
   \includegraphics[width=0.48\textwidth]{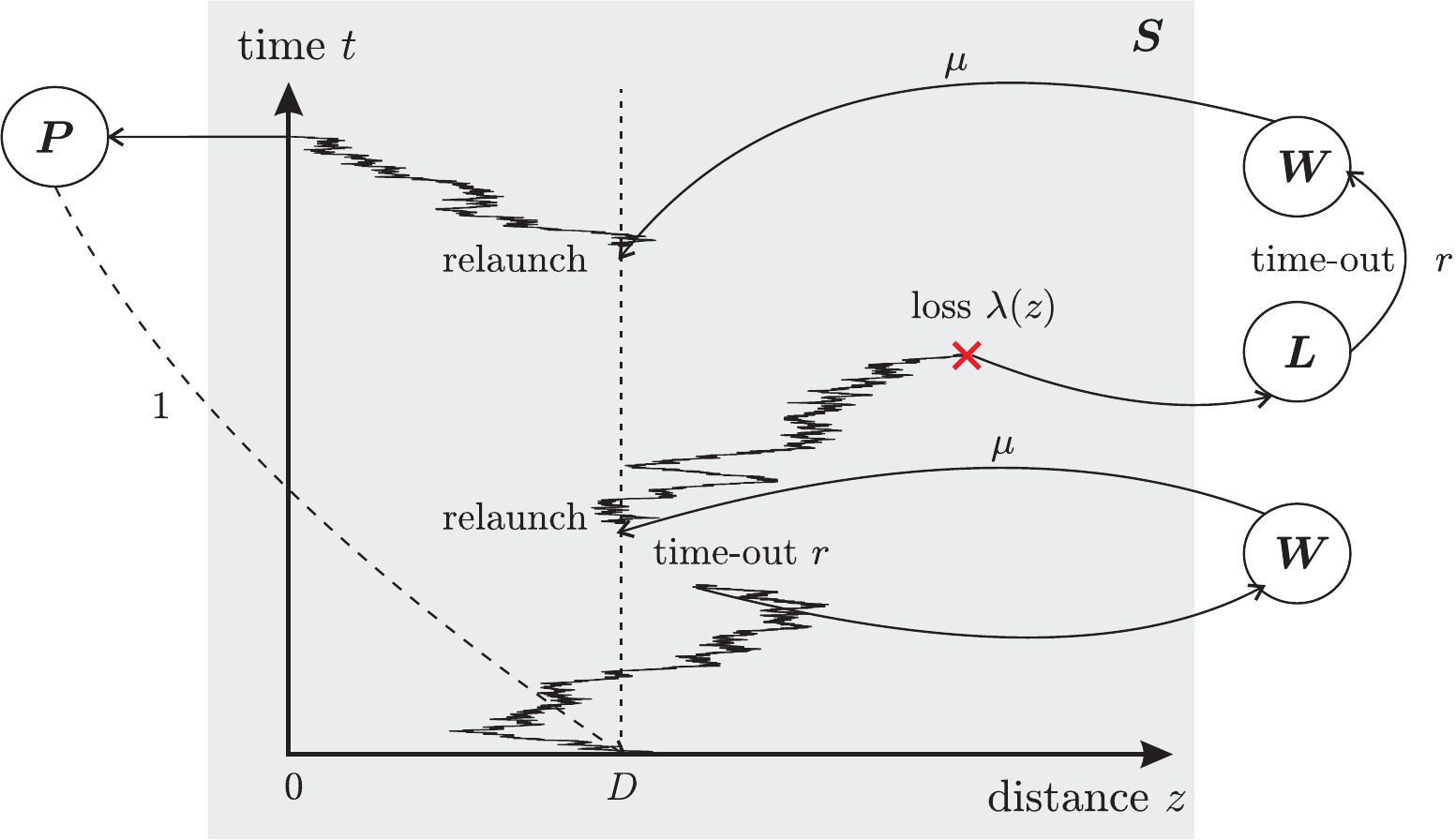}
   \caption{The search process with $N=1$: in this example the search is relaunched twice due to timeout and loss, before the object is finally found. Then, after one time unit, the search process starts again as before at the source with a new searcher being sent out.} \label{diagram}
\end{figure}

A schematic representation of the search process for $N=1$ is presented in Figure~\ref{diagram}. The state of the $i$-th searcher at time $t\geq 0$ is $s_i(t)$ which can take one of the values $\{{\bf S_i,L_i,W_i,P_i}\}$ defined as follows:
\begin{itemize}
\item $\bf S_i$: If the $i$-th searcher is searching and its distance from the destination is $Z_i(t)>0$. We denote the probability density function (pdf) of the distance $Z_i(t)$ by $f_i(z_i,t)dz_i=P[z_i<Z_i(t)\leq z_i+dz_i, ~s_i(t)={\bf S_i}]$.

\item $\bf W_i$: The $i$-th searcher's life-span has ended, and so has its search. Note that this may have happened because it was destroyed or became lost, but this becomes known to the source via the time-out which is exponentially distributed with parameter $r$. After an additional exponentially distributed delay of parameter $\mu$, it is replaced at the source by a new searcher with the same identity.

\item $\bf L_i$: The $i$-th searcher has been destroyed or lost, and its search is ended; for small $\Delta t$ and $Z_i(t)=z>0$, this happens with a probability $\lambda(z) \Delta t + o(\Delta t)$, where $\lambda(z)\geq 0$ is the loss rate at distance $z$. The time spent in this state is exponentially distributed with the same parameter $r$ as the life-span since the source realises that the searcher is lost or destroyed via the time-out effect. At the end of this exponentially distributed time, the searcher is handled just as if it has ``died''.
\item $\bf P_i$: The $i$-th searcher has found the object being sought, and its search process stops.
\end{itemize}

$\{Z_i(t):t\geq 0\}$ is modelled as a diffusion process, and when $Z_i(t)=z$:
\begin{itemize}
\item The average change in the searcher's distance to the object being sought in a small time interval $\Delta t$ is $b_i(z) \Delta t$ and $b_i(z)\approx \lim_{\Delta t\rightarrow 0}\frac{E[Z_i(t+\Delta t)-Z_i(t) |Z_i(t)=z]}{\Delta t}$,
\item The variance of the distance travelled by the searcher over the same time interval is $c_i(z) \Delta t$, where $c_i(z)\approx \lim_{\Delta t\rightarrow 0}\frac{E[(Z_i(t+\Delta t)-Z_i(t))^2]-(E[Z_i(t+\Delta t)-Z_i(t)])^2| Z_i(t)=z]}{\Delta t}$.
\end{itemize}
However, for $N>1$ we also need to represent the ``race'' between the $N$ searchers, and the interaction between them due to the fact that when the first one reaches the object, the progress of all others is stopped. This race is modelled in \cite{Gelenbe2010} using a parameter $a_i(t)$, $1\leq i\leq N$, which represents the total rate of attraction exerted at time $t$ on the $i$-th searcher, by all other search processes, due to the fact that one of the others may have finished its search.

The ergodic process $\{s_i(t): t\geq 0\}$ can then be expressed in terms of a system of coupled equations describing a mixed continuous space (diffusion) supplemented by a discrete Markov process, where the discrete part describes the states in which the search agents finds itself when it is not actually searching. The searcher can enter the rest state $\bf P_i$ from the ``active'' state $\bf S_i$, the ``lost'' state $\bf L_i$ and the ``time-out before retransmission'' state $\bf W_i$. We also see in Figure~\ref{diagram} that the searcher can enter the lost state from any position $z_i>0$, and that a time-out can occur for a searcher that is in the lost state. Since the behaviour of all searchers when they are not in the rest state are independent, it follows that the event that triggers the jump of searcher $i$ into the rest state does not depend on the prior state of searcher $i$ but on the state of the {\em other} searchers.
Obviously the case with $b_i(z)<0$ is the most favourable, since the searcher is getting on average closer to the destination with time. When  $b_i(z)>0$,  the searcher on average moves {\em away} from the object of interest, for instance because intermediate locations provide wrong information on average. When the searcher lacks information altogether
and is on average not getting closer nor further from the object, then $b_i(z)=0$.

In most applications including web search, both the time and energy needed for a successful search are of interest, and we may assume that when the search agent is actually searching, i.e.,
when it is in state $\bf S_i$, it consumes energy at a rate of one energy unit per unit time
due to the use of computational processors and access to memories.
No energy is being consumed by the  search agent when it has been disabled or when the source is waiting to send out another searcher.

$J_{k,N}^-$ will denote the total energy consumed by the $N$ searchers from time $t=0$ up to the instant when
the minimum
required number of $k$ search agents have found the object. In the best of cases all remaining $N-k$ search agents can then be stopped simultaneously.

However we may not be able to stop the remaining search agents
through a common control program. In that case they will stop later either because they successfully complete, or are destroyed or lost, or they are stopped by time-outs. In this case we denote by  $J_{k,N}^+$ the total energy that is consumed.

\section{Answering the Basic Questions}\label{avg-comput}

Consider the most common case where the search is successful as soon as any {\em one} of the $N$ searcher agents actually finds the object.
To do this we construct an
indefinitely repeating ergodic search process which simulates a situation in which all the $N$ searchers are sent out at time $t=0$ and as soon as the first of these  finds the object, all other searchers are artificially stopped and instantaneously moved to the rest state where they remain for one time unit. After this ``rest''  the process is repeated indefinitely in the same manner, and state $\bf P$ becomes a synchronised re-start state for all of the searchers.

This transformation of the initial problem of finding the search time now allows us to use the steady-state distributions of the recurrent process so as to compute the average time $E[T_{1,N}]$ needed from any successive start of the search until the first instance when state $\bf P$ is reached again. If we denote by $P$ the steady-state probability that the recurrent process we have just described is in state $\bf P$, then it is easy to see that $E[T_{1,N}] = P^{-1}-1$.

To answer Question 1 of Section \ref{Basic}, when the medium in which the searchers move is homogeneous, i.e. $b(z)=b$, $c(z)=c$ and $\lambda(z)=\lambda$,  the
the average time $E[T_{1,N}]$ for
the most successful searcher to find the object \cite{Gelenbe2010} includes
the searcher's loss rate $\lambda$, the average drift $b$ so that $b<0$ means that on average at each step the searcher gets closer to the object being sought,
$c\geq 0$ which is the second moment parameter of the searcher's random motion per unit time, while $r^{-1}$ is the average time-out used by the sender at the source, $\mu^{-1}$ is the average additional wait time before the search restarts, and $a$ is computed numerically as indicated below:
\begin{equation}\label{time}
E[T_{1,N}]  = \frac{\mu+r+a}{N(\mu+a)(r+a)} [e^{\frac{D}{c}[b+\sqrt{b^2+2c(\lambda+r+a)}]}-1]
\end{equation}
where $a=(N-1)/\{N[1+E[T_{1,N}]\}$ so that $a=0$ for $N=1$.

From this expression we make further observations,
and answer Question 2 of Section \ref{Basic}:
\begin{itemize}
\item If the search is deterministic so that $c=0$, then a simple calculation
using L'H\^{o}pital's rule tells us that the search time is finite if $b<0$ and the searcher always heads towards the object being sought,
while the object will never be found if $b\geq 0$.
\item On the other hand:
\begin{itemize}
\item If the search process is randomised so that $c>0$  and either $\lambda >0$ and there are losses, or $r >0$ and there is a finite time-out, surprisingly enough the search time is {\em finite}
\underline{even} when at each step the searcher gets {\em on average further away} from the object being sought, i.e. $b\geq 0$. This is due to the fact that the
loss with $\lambda>0$ or the time-out with $r >0$, both act to curtail a search which is being unsuccessful, and the process is repeatedly restarted after the time-out
until a given search is successful.
\item The search time is finite also in the obvious case when the searcher gets on average closer to the object being sought $b<0$.
\end{itemize}
\end{itemize}

Let us now turn to Question 3 of \ref{Basic}. Since energy is consumed by the search as a unit of energy per unit of search time when the
search agents move through the search space, and hence {\em are not} resting in one of the states ${\bf L_i,~W_i,~P}$, the energy expended by the search is proportional to
the number of the searchers $N$ and to the time they spend in the state ${\bf S_i}$, namely
$E[J_{1,N}^-] = N(1+E[T_{1,N}])\int_0^\infty f_i(z)dz$, since all searchers are statistically identical.

This allows us to use \eqref{time} to compute the average amount of energy which is expended by all of the $N$ searchers together until the instant when the object is found:
\begin{equation}
E[J_{1,N}^-] = \frac{1}{\lambda + r + a} [e^{\frac{D}{c}[b+\sqrt{b^2+2c(\lambda+r+a)}]}-1] \label{energy}
\end{equation}

\begin{figure}[]\centering
\includegraphics[width = 0.38\textwidth]{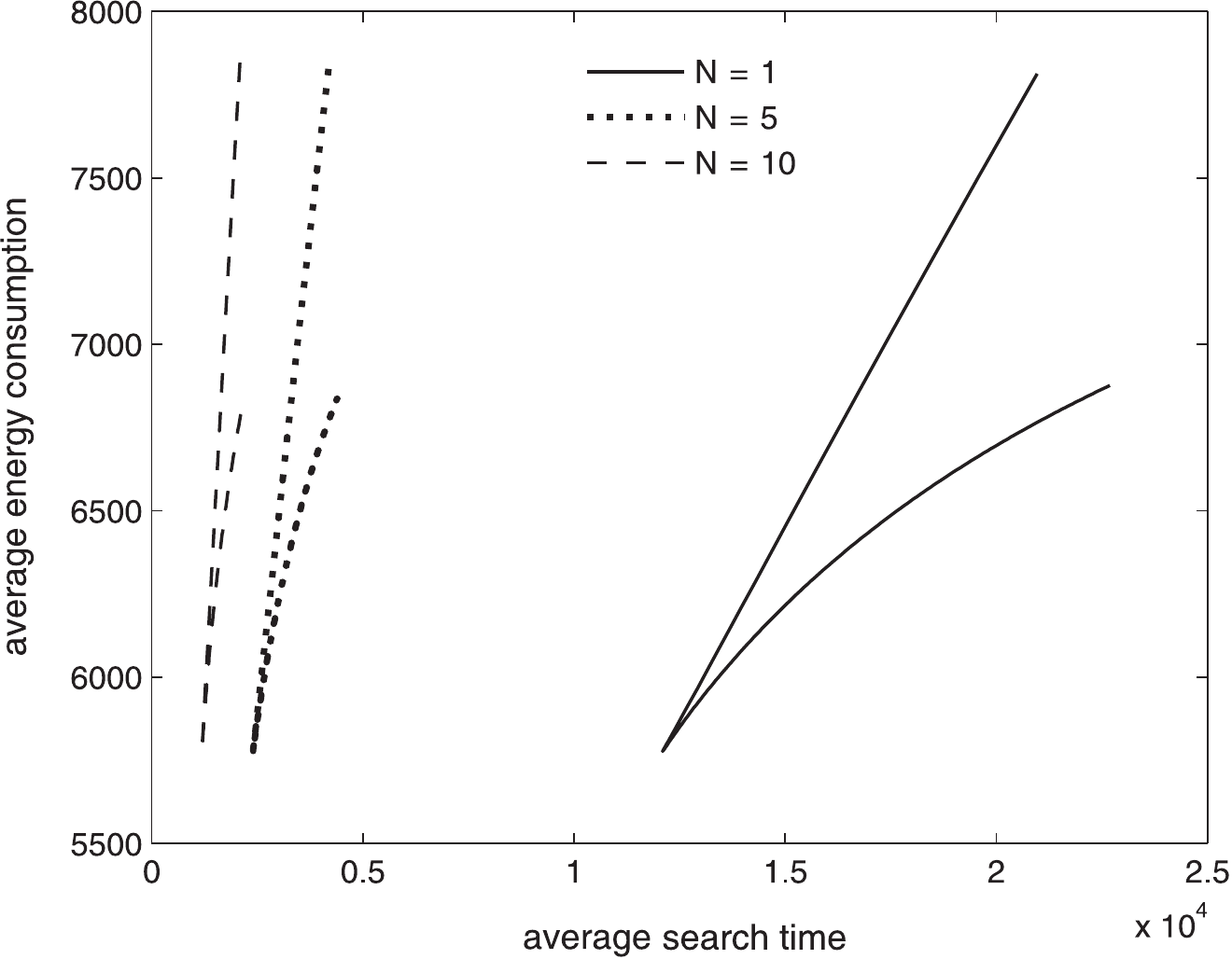} \\\vspace{0.2in}
\includegraphics[width = 0.38\textwidth]{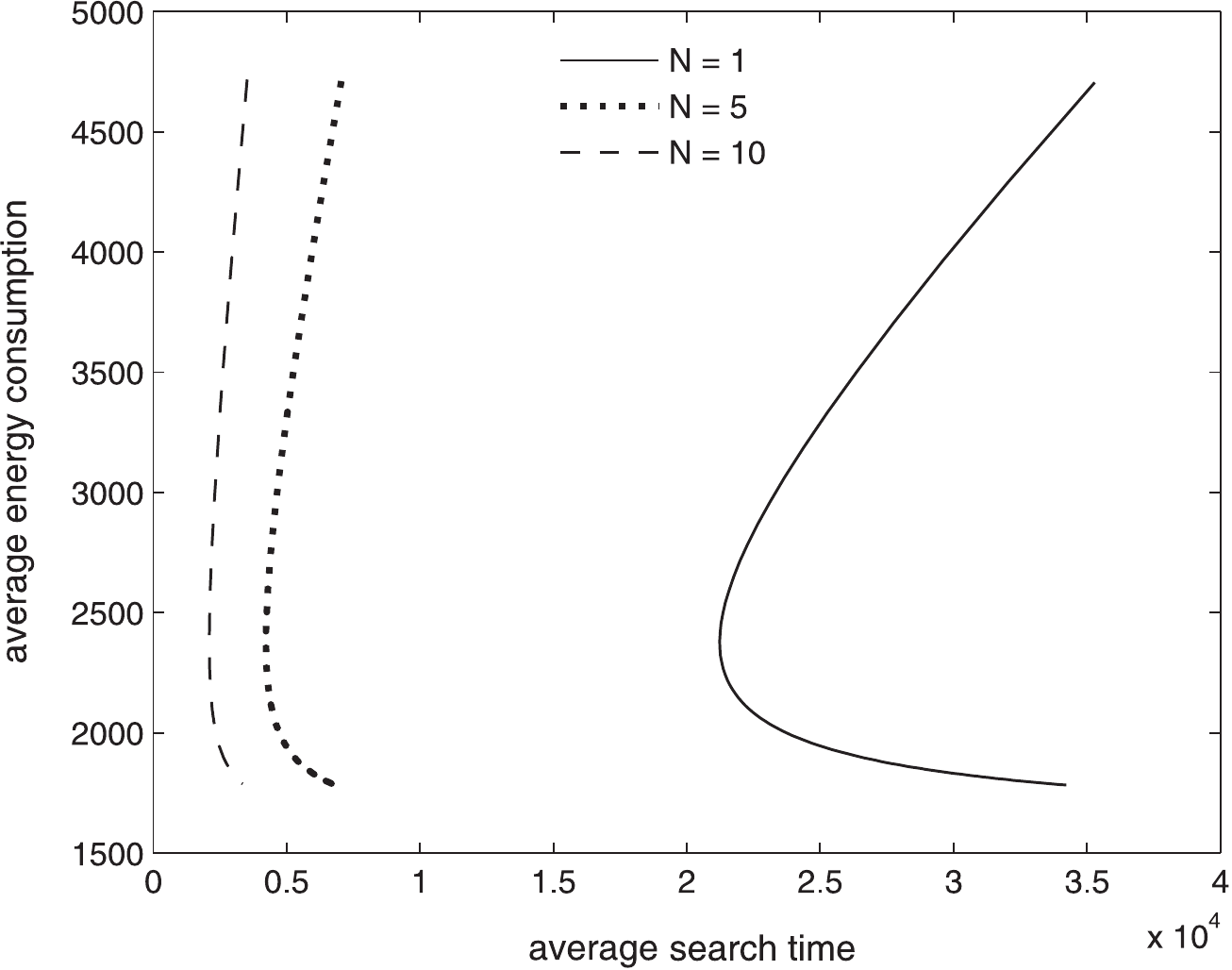}
\caption{The locus of $E[T_{1,N}]$ and $E[J_{1,N}^-]$ when the average time-out $1/r$ is varied. The parameters are $D = 10, c=1, \mu=0.05$ and (a) $b=0.2, \lambda=0.01$; (b) $b=0,\lambda=0.15$. For low loss rate and high uncertainty in search direction, the minimum energy consumption is obtained when the average search time is also minimum, while with high loss rate, minimum search time does not yield minimum energy consumption.}\label{fig-tradeoff}
\end{figure}

In order to see how $N$ should be chosen to optimise both search time and energy, we have plotted the locus of $E[T_{1,N}]$ and $E[J_{1,N}^-]$ when the average time-out $1/r$ is varied for $b=0.2$ and $\lambda=0.01$ in Figure~\ref{fig-tradeoff}(a). The relatively unfavourable case when the searcher does not really know which direction it should search,
with  $b=0$ is shown in Figure~\ref{fig-tradeoff}(b).

In both of these cases there is definitely a range of values
of $1/r$ where both energy and time are at  close to their minima.

\subsection{Energy consumption}\label{energy}

\begin{figure}[t]\centering
   \includegraphics[width=0.42\textwidth]{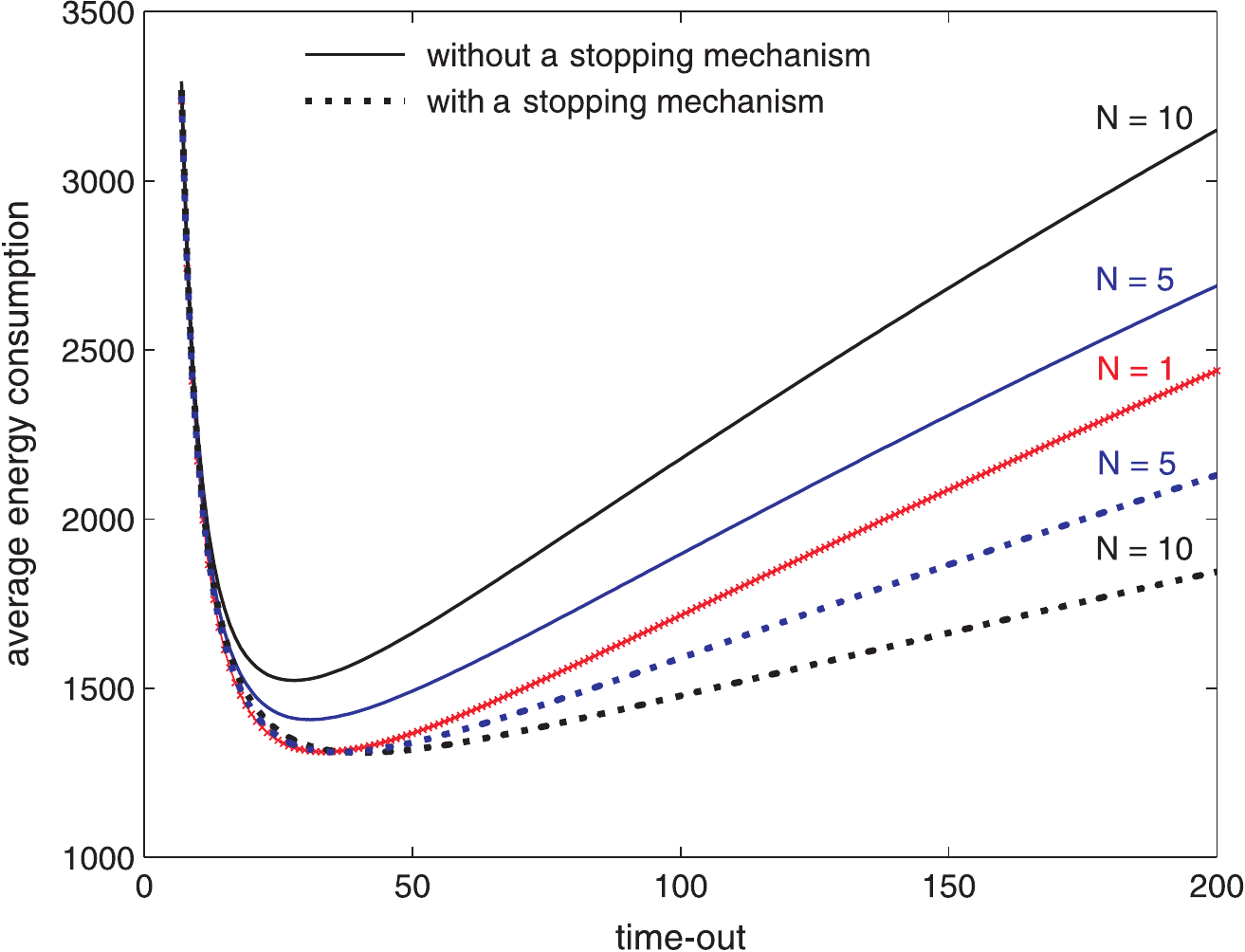}
   \caption{Average energy consumption (with and without a stopping mechanism) versus time-out $1/r$ for $k=1, b=0.15, c=1.25, \lambda=0.001, \mu=0.1, D=10$ and different values of $N$. When $N$ increases, the {\em energy consumed increases if there is no stopping mechanism}, and the opposite is true with the stopping mechanism.}\label{timeout}
\end{figure}

\begin{figure}\centering
   \includegraphics[width=0.42\textwidth]{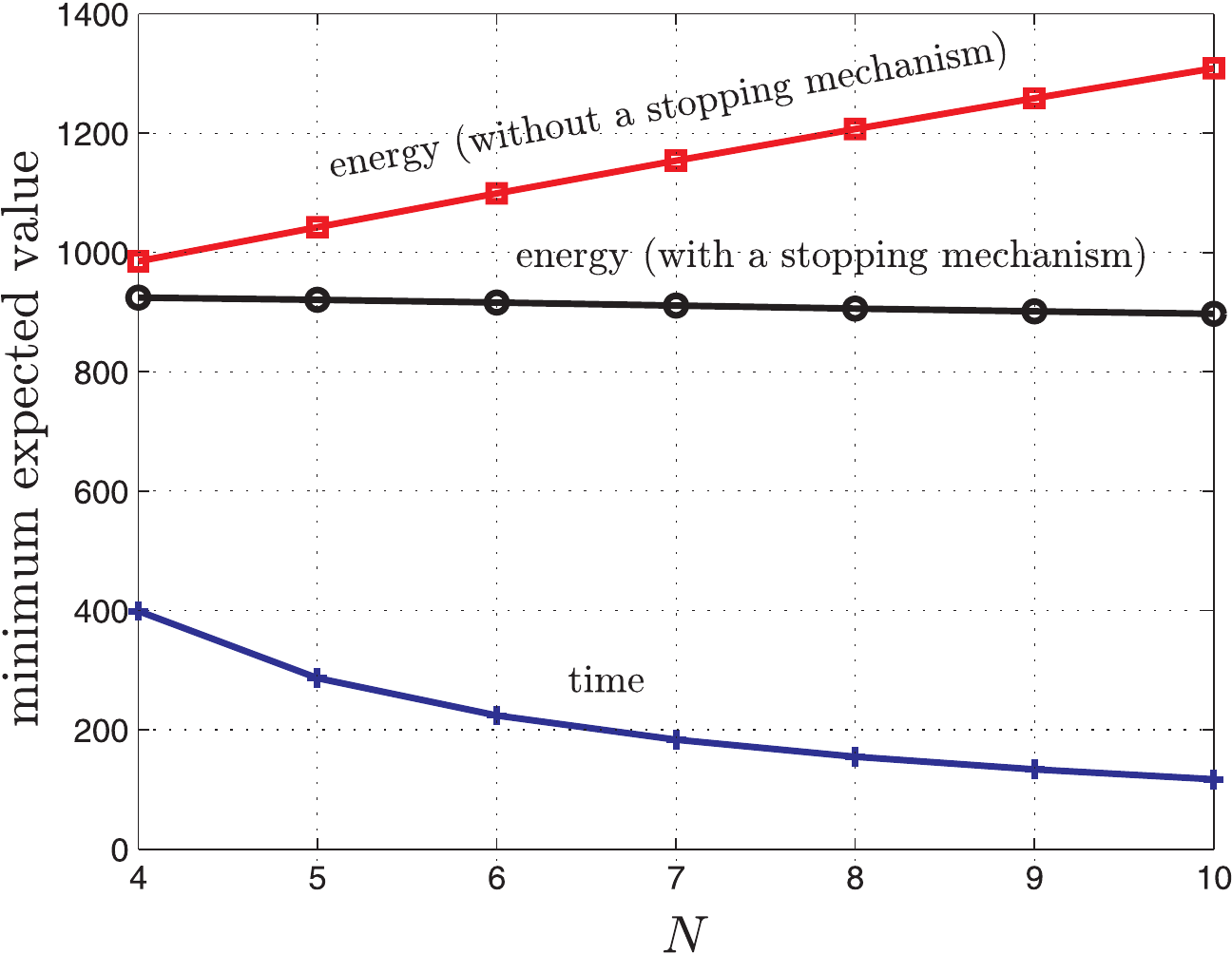}\\\vspace{0.2in}
   \includegraphics[width=0.42\textwidth]{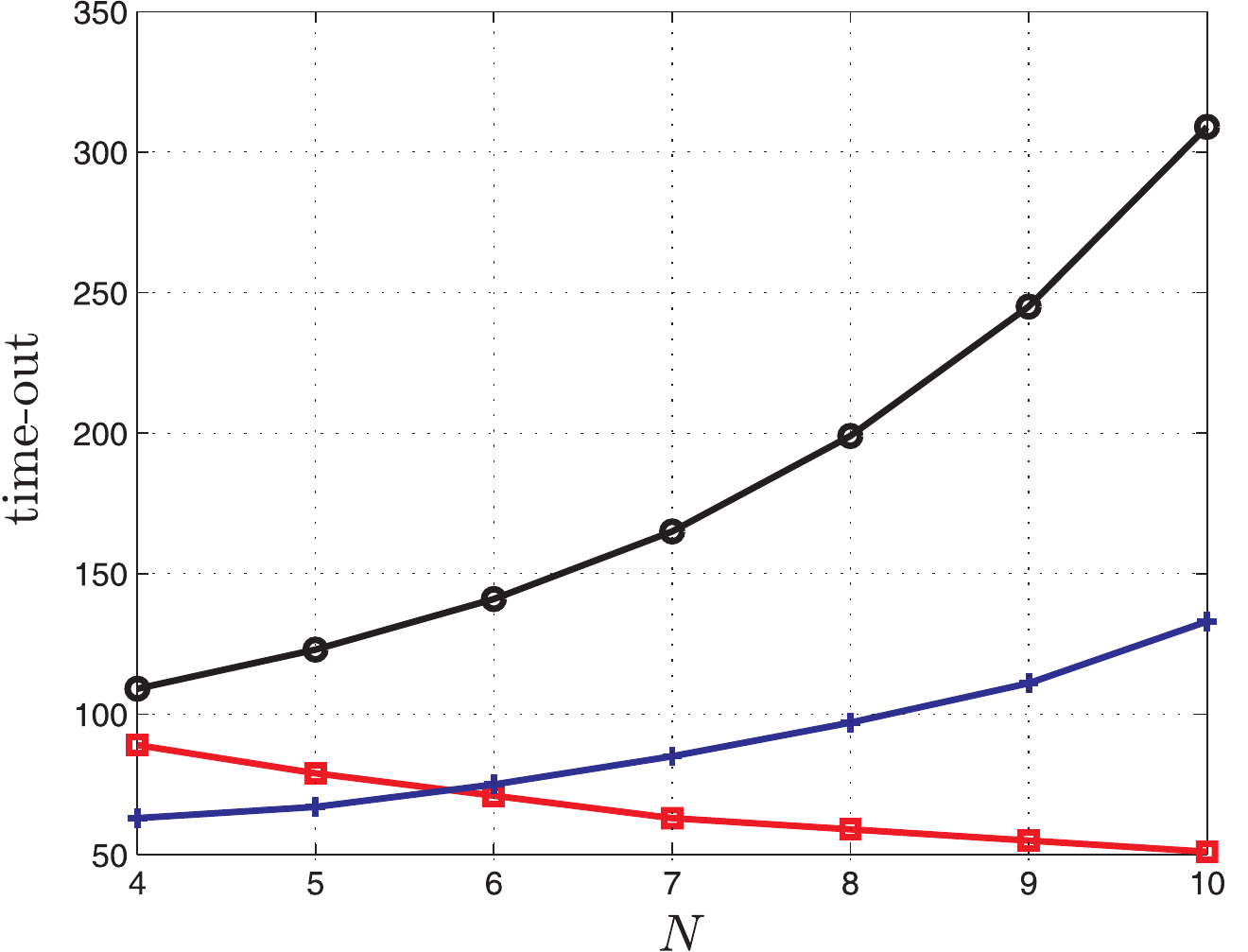}
   \caption{(a) Minimum achievable average search time and energy and (b) the corresponding time-out, versus $N$ for $k=3, b=0, c=1, \lambda=0.0025, \mu=0.1, D=10$}\label{Opt}
\end{figure}

Figure~\ref{timeout} shows the average energy consumed by the $N$ searchers if they are all stopped as soon as $k$ are successful: we see that the energy consumed decreases as $N$ increases. However, when the successful searchers stop but the unsuccessful ones continue until a time-out or until they are destroyed by some other cause, we observe that the energy consumed increases with $N$. In Figure~\ref{Opt}(a) we plot the minimum achievable average search time and energy consumption versus the number of searchers $N$, and Figure~\ref{Opt}(b) shows the corresponding optimum time-outs. One sees that the minimum energy consumed until the object is found (i.e. $E[J_{k,N}^-]$) does not vary much with the number of searchers $N$. However, in the absence of a stopping mechanism the minimum energy consumed by the search $E[J_{k,N}^+]$ {\em increases} with $N$ while the optimum time-out decreases in order to reduce the additional energy wasted by active searchers after the completion of the search

\section{A Simple Asymptotic Formula}  \label{asymptot1}

Consider a single searcher, and denote by $T$ and $J$ its search time and energy consumption, respectively. The probability that the searcher has reached the object by time $t$ is $G(t)\equiv \Pr[T\leq t]$ and its pdf $g(t)$.

If the searcher is successful in locating the object in its first attempt then the search time and energy consumption are equivalent. On the other hand, if the search is interrupted at least once then $T$ will exceed $J$ by the amount of time spent in states $\mathbf{L}$ and $\mathbf{W}$ waiting for the search to be relaunched.

Therefore, the joint pdf of $T$ and $J$ can be obtained by accounting for the possibilities of locating the object in $1,2,\cdots$ attempts while including the time spent in states $\mathbf{L}$ and $\mathbf{W}$ in $T$ but not in $J$.

Note that an attempt to locate the object can be interrupted by either time-out or loss, which are mutually independent and exponentially distributed random variables with parameters $r$ and $\lambda$. Thus the pdf of the duration of a search time until its first interruption is $(\lambda+r)e^{-(\lambda+r)t}[1-G_0(t)]$, where $G_0(t)$ is the probability that a pure diffusion process starting at distance $D$ reaches destination by time $t$.

The search may be interrupted randomly several times in this manner, and after each interruption it starts again at the source after a further delay. The last and hence successful attempt at reaching the destination has a duration whose pdf is $g_0(t)e^{-(\lambda+r)t}$. Since each attempt is independent of its predecessors, it is then straightforward to compute the Laplace transform (LT) of the joint density of $T$ and $J$.

The probability that $k$ out of $N$ independent searchers will be successful by time $t$ is $G_{k,N}(t)\equiv \Pr[T_{k,N}\leq t] =\binom{N}{k}G(t)^k[1-G(t)]^{N-k}$. Define $G^{-1}(p) = \inf\{t:~G(t) \geq p\},~ 0<p<1$, the {\em quantile function} of the distribution of the search time for a single searcher. When $N$ is large, it is known that $T_{\lceil pN \rceil,N}$, the $p$-th sample quantile, is asymptotically normally distributed: $T_{\lceil pN \rceil,N} \sim \mathcal{N}\left(G^{-1}(p),\frac{p(1-p)}{N~[g(G^{-1}(p))]^2} \right)$.

\begin{figure}[t]\centering
   \includegraphics[width=0.42\textwidth]{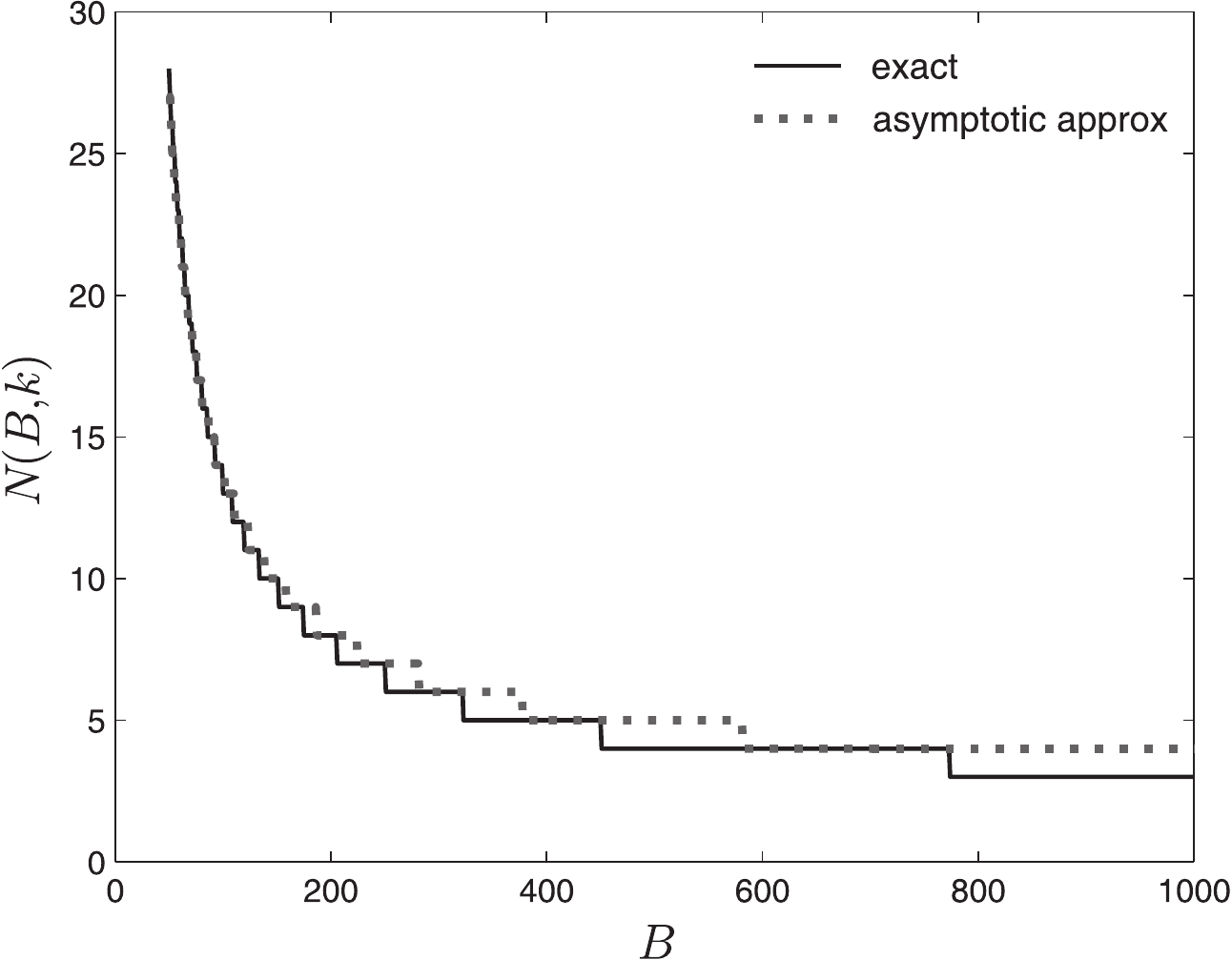}
   \caption{Comparison of the asymptotic approximation with exact analysis for the total number of searchers $N(B,k)$ that are required so that $k=3$ of them find the object within time $B$. Here $b=0$, $c=1$, $\lambda=0.0025$, $r^{-1} = 78$, $\mu^{-1}=10$ and $D=10$. } \label{noSearchers}
\end{figure}

Thus for large $N$ the distribution of the time for $k$ out of $N$ searchers to be successful tends to a constant equal to the $p\approx k/N$-th quantile of $G(t)$. As a consequence, the number of searchers $N(B,k)$ required to find the object in time $B$ when $N$ is large is given approximately by:
\begin{equation}
N(B,k) \cong \left\lceil \frac{k}{G(B)} \right\rceil \label{asymp}
\end{equation}
Since convergence to the normal distribution is fast, the expression \eqref{asymp} provides a good approximation even for relatively small $N(B,k)$. The good agreement between the asymptotic approximation of \eqref{asymp} and the
detailed analysis for $G_{k,N}(B)$ is illustrated in Figure~\ref{noSearchers}.

\section{Search in a Non-homogeneous Medium}\label{non-hom}

A non-homogeneous search space may be  motivated
by the case where the object being sought is protected and well hidden: as the searcher approaches, its progress becomes more frequently blocked or destroyed, and a new searcher has to be sent out to replace it. However the search may progress faster as the searcher approaches the object
since it may now have better information.

This case can be analysed by using a finite number of ``segments'' to represent the search space with  different parameters describing the searcher's movement as a function of its distance to the object:\\
\\
 $ \includegraphics[width=0.49 \textwidth]{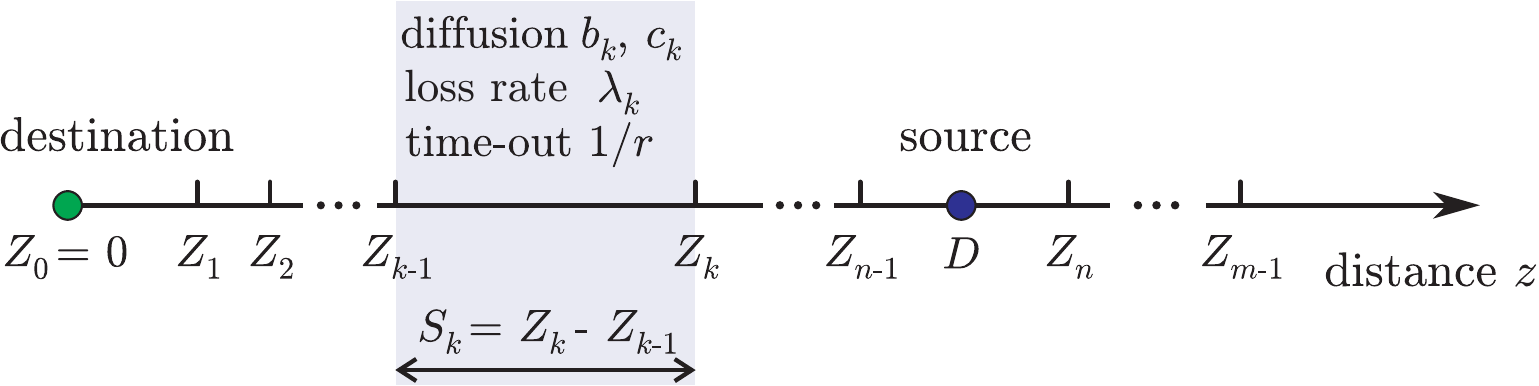}$\\
The first segment is in the immediate proximity of the object starting at distance $z=0$. Each segment may have a different size, and  as many segments as are needed can be used to describe the search accurately, with each segment being as small as necessary all not necessarily of the same size. This leads to a neat algebraic ``product form'' representation for the average search time \cite{CJ-diffusion2012}.

\begin{figure}[t]\centering
\includegraphics[width=0.42\textwidth]{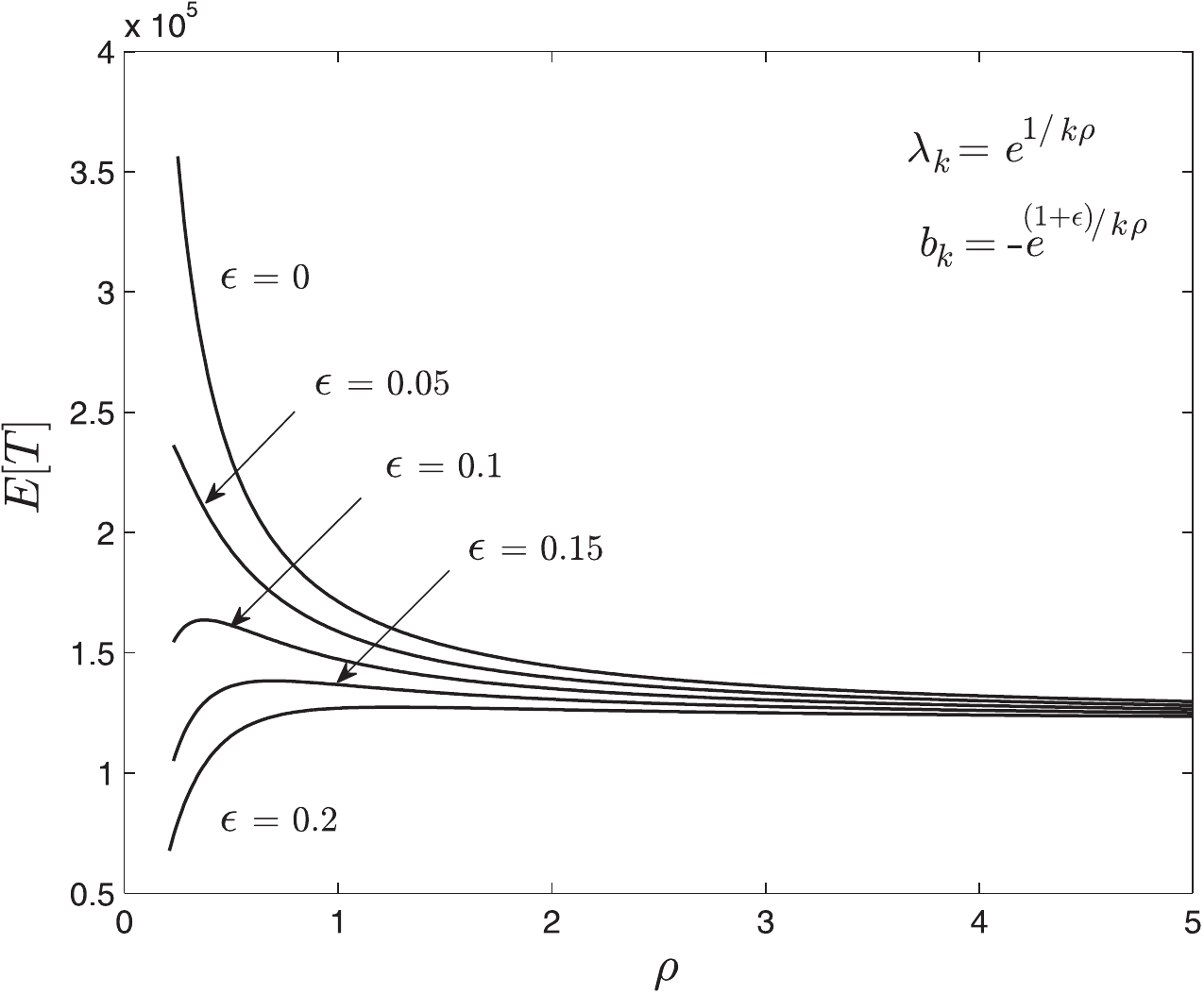}
\caption{ \label{fig7} Average search time $E[T]$ versus $\rho$ when $\lambda_k = e^{\frac{1}{k\rho}}$ and $b_k = - e^{\frac{1+\epsilon}{k\rho}}$ for different values of $\epsilon\geq 0$; $c_k=1$, $D=10$, $r=0.05$, $\mu=0.025$ and $S_k = 1$ for $k < m=20$.}
\end{figure}

The destruction of the searcher followed by a time-out, as well as just the time-out itself,  relaunches the search process allowing the searcher to improve its chance to find the object. However if the object being sought is well hidden then the searcher may never attain the object. Figure~\ref{fig7} shows for successive segments $k$ of the search space, when the searcher's destruction rate (defence) varies as  $\lambda_k\approx e^{\frac{1}{k\rho}}$ while the rate of approaching the object improves
as  $b_k \approx -e^{\frac{1+\epsilon}{k\rho}}$ with $\epsilon \geq 0$, then the average time $E[T]$ to find the object tends to infinity even when near the object the search speed is greater and its randomness is smaller. However if  the searcher's speed of approach to the object grows faster than the rate at which the searcher is  destroyed, then $E[T]$ remains finite and can tend to zero, while in the opposite case it will tend to infinity.

\section{Other Topics of Interest}

Many other aspects of search are not addressed in this paper. Load sharing or balancing so as to achieve overall better performance in search is of great interest, to subdivide work among multiple subsets of searchers. It is also interesting to exploit the distinct capacities of multiple classes of searchers that are specialised in different features so as to improve both the time and energy needed for a search.
Another interesting area concerns how agents learn from each other. Searchers may also  conceal their knowledge of the location of an object, or deceive other searchers so as to impede them and maximise their own chances of success. Malicious agents may use viruses \cite{Viruses} to infect some other agents, and must stop their search from being successful, so that one may study schemes that optimally impede the search, rather than make it easier.
Thus this area still reserves many opportunities to study problems that are of value to search in big data sets and networks.


\begin{biographynophoto}{Erol Gelenbe}
is the Denis Gabor Chair Professor at Imperial College in the Department of Electrical and Electronic Engineering. A Fellow of IEEE and ACM, he graduated from the Middle East Technical University (Ankara) and holds MSc and PhD degrees from the Polytechnic Institute of New York University. A Member of the French National Academy of Engineering and of the Science Academies of Hungary, Poland and Turkey, he is known for inventing the G-Network queueing network models and the Random Neural Network Model and deriving their product form solutions, he is a designer of the QNAP performance modeling package and of the FLEXSIM object-oriented Flexible Manufacturing System simulation method, and developed the first random access fiber optics network (XANTHOS), the multiprocessor voice packet based switch SYCOMORE and the autonomic software defined network protocol CPN.  He has received several prizes including ACM SIGMETRICS' Life-Time Achievement Award, the IET Oliver Lodge Medal, the {\em Grand Prix France Telecom}, and the {\em In Memoriam Dennis Gabor Award} from Hungary.
\end{biographynophoto}

\begin{biographynophoto}{Omer H. Abdelrahman}
is a Research Associate in the Intelligent Systems and Networks Group at Imperial College with a BSc in electrical engineering from University of Khartoum, Sudan. He holds an MSc in communications and signal processing, and a PhD in computer networks, both from Imperial College. His research interests include stochastic performance models, search techniques in uncertain environments, and network security, and he is active in the EU FP7 Project NEMESYS.
\end{biographynophoto}

\end{document}